\documentstyle[editedvolume,namedreferences,epsf]{crckapb} 

\def\ga{\mathrel{\mathchoice {\vcenter{\offinterlineskip\halign{\hfil
$\displaystyle##$\hfil\cr>\cr\sim\cr}}}
{\vcenter{\offinterlineskip\halign{\hfil$\textstyle##$\hfil\cr  
>\cr\sim\cr}}}
{\vcenter{\offinterlineskip\halign{\hfil$\scriptstyle##$\hfil\cr
>\cr\sim\cr}}}  
{\vcenter{\offinterlineskip\halign{\hfil$\scriptscriptstyle##$\hfil\cr
>\cr\sim\cr}}}}}
\def\la{\mathrel{\mathchoice {\vcenter{\offinterlineskip\halign{\hfil
$\displaystyle##$\hfil\cr<\cr\sim\cr}}}
{\vcenter{\offinterlineskip\halign{\hfil$\textstyle##$\hfil\cr
<\cr\sim\cr}}}
{\vcenter{\offinterlineskip\halign{\hfil$\scriptstyle##$\hfil\cr
<\cr\sim\cr}}}
{\vcenter{\offinterlineskip\halign{\hfil$\scriptscriptstyle##$\hfil\cr
<\cr\sim\cr}}}}}
\begin{opening}
\title{\hbox{EVOLUTION ON THE AGB AND BEYOND: ON THE} 
FORMATION OF H-DEFICIENT POST-AGB STARS}

\author{THOMAS BL\"OCKER}
\institute{Max-Planck-Institut f\"ur Radioastronomie, 
Bonn, Germany}

\end{opening}

\runningtitle{EVOLUTION ON THE AGB AND BEYOND}

\begin{document}
\begin{abstract}
The evolution on the AGB and beyond is reviewed with respect to the origin  
of Wolf-Rayet central stars. 
We focus on thermal pulses due to their 
particular importance for the evolution of hydrogen deficient stars. 
It is shown that overshoot applied to all convection regions is a key 
ingredient to model these objects leading to
intershell abundances already close to the surface abundances of Wolf-Rayet
central stars. In contrast to standard evolutionary calculations, 
overshoot models do show dredge up for very low envelope masses 
and efficient dredge up was found even during the post-AGB 
stage. Three thermal pulse scenarios for Wolf-Rayet central stars can now
be distinguished: an AGB Final Thermal Pulse (AFTP) occurring at the very end 
of the AGB evolution, a Late Thermal Pulse (LTP) occurring during the post-AGB
evolution when  hydrogen burning is still on, and a Very  Late Thermal Pulse 
(VLTP)  occurring on the cooling branch when hydrogen burning has already 
ceased. All scenarios lead to hydrogen-deficient post-AGB stars with 
carbon and oxygen abundances as observed for Wolf-Rayet stars.
Hydrogen is either diluted by dredge up (AFTP, LTP) or completely 
burnt (VLTP). 
\end{abstract}

\section{Introduction}
The origin of hydrogen-deficient post-Asymptotic-Giant-Branch (post-AGB)
stars has been a matter of debate since many years. Although stars
evolving through the AGB phase stay hydrogen-rich at their
surfaces, a considerable fraction of their descendants, the
central stars of planetary nebulae (CSPNe),
show hydrogen-deficient compositions (Mendez 1991).
Approximately 20\% of the whole CSPNe population
appears to be hydrogen deficient
while the rest show solar-like compositions.
Whereas we believe to understand the structure and evolution of the latter
with normal surface abundances reasonably well, the evolution of the former
appeared to be still enigmatic.
Important constituents of the hydrogen-deficient population
are the so-called Wolf-Rayet (WR) central stars and the hot
PG\,1159 stars with typical surface abundances of (He,C,O)=(33,50,17)
by mass (Dreizler \& Heber 1998, Koesterke \& Hamann 1997).
In addition to their exotic abundances, WR CSPNe expose strong stellar winds
of $\sim 10^{-6} $M$_{\odot}$/yr and higher PN outflow velocities but 
otherwise do not differ significantly from the rest of the central stars
(Gorny \& Stasinska 1995, Tylenda 1996) and are 
expected to have the same mean mass of 
$\sim 0.6 $M$_{\odot}$ (Szczerba et al. 1998). 
Standard stellar evolution calculations
predict post-AGB stars either to have hydrogen-rich surfaces 
(e.g.\ Vassiliadis \& Wood 1994, Bl\"ocker 1995b)
or, if hydrogen-deficient, to expose only a few percent of oxygen 
in their photospheres (Iben \& McDonald 1995), and thus appropriate models
for WR CSPNe are still lacking.

Three types of evolutionary scenarios have been invoked to explain
the Wolf-Rayet central stars. The first assumes the Wolf-Rayet stars
to be the products of binary evolution with a common envelope phase
(Tylenda \& Gorny 1993) but appears to be unlikely because binarity is
not observed. The other scenarios assume that they are either formed
by a final thermal pulse during their post-AGB evolution or directly emerge
from the AGB.

\section{Evolution along the AGB}
Stars with initial masses between 0.8 and 6--$8 M_{\odot}$
evolve after completion of central hydrogen and helium burning
through the AGB phase consisting then of a compact carbon/oxygen core
($\sim 0.5 ... 1 M_{\odot}$) surrounded by a thin layer 
of some $10^{-2}$\,M$_{\odot}$ occupied by two shells burning helium and
hydrogen, resp., and a huge, almost fully convective envelope.
Although being short the AGB phase is of essential importance since it is
governed by a rich nucleosynthesis, various processes which
mix up processed material from the interior to the surface, and strong
stellar winds, which substanially erode the stellar surfaces thereby
continuously enriching the interstellar medium with heavier elements.
This concerns in particular the evolution on the upper AGB
which is dominated by continuously increasing mass loss
(see Habing (1996) for a review).
Observations indicate rates of $10^{-7}$\,M$_{\odot}$/yr for small-period
Mira stars and up to $10^{-4}$\,M$_{\odot}$/yr for luminous 
long-period variables (Wood 1997). These winds are most likely
driven by dust and shocks (Winters 1998) 
leading at larger mass-loss rates to the complete obscuration of the star
by a dusty circumstellar envelope.

Concerning mixing processes and nucleosynthesis,
thermal instabilities of the helium burning shell
(thermal pulses) and the possible penetration of the convective
envelope into the hydrogen burning shell (hot bottom burning, HBB) 
are crucial for the evolution on the upper AGB.

Hot bottom burning is restricted to more massive stars
($M \ga 4 $M$_{\odot}$). Temperatures in excess of $50 \cdot 10^{6}$\,K
can be  reached at the base of the convective envelope and material burnt there
is immediately mixed to the surface leading, for instance, to the formation
of lithium-rich AGB stars (Scalo et al.\ 1975).
HBB  models do not obey Paczynski's (1970)
classical core-mass luminosity relation but, instead, evolve rapidly
to very high luminosities (Bl\"ocker \& Sch\"onberner 1991). Due to 
CN cycling of the envelope $^{12}{\rm C}$ can be transformed into
$^{13}{\rm C}$ and $^{14}{\rm N}$. Consequently, a low
$^{12}$C/$^{13}$C ratio is a typical signature of HBB
that can delay or even prevent
AGB stars from becoming carbon stars (Iben 1975; Boothroyd et al.\ 1993).
Finally, when mass loss has substantially reduced the envelope mass, 
the envelope convection moves upwards again and
HBB shuts down (see Bl\"ocker 1995a).
Planetary nebulae showing abundance patterns with  the signature of former HBB
are discussed in, e.g.\ Kaler \& Jacoby (1989) and Clegg (1991).

In the following we want to focus on thermal pulses due to their particular
relevance for the origin and evolution of Wolf-Rayet central stars.
A more general review of the AGB evolution is given in Bl\"ocker (1999).
%

\subsection{Thermal pulses and third dredge up}
\begin{figure}
\centering
\epsfxsize=0.8\textwidth
\mbox{\epsffile[70 30 570 440]{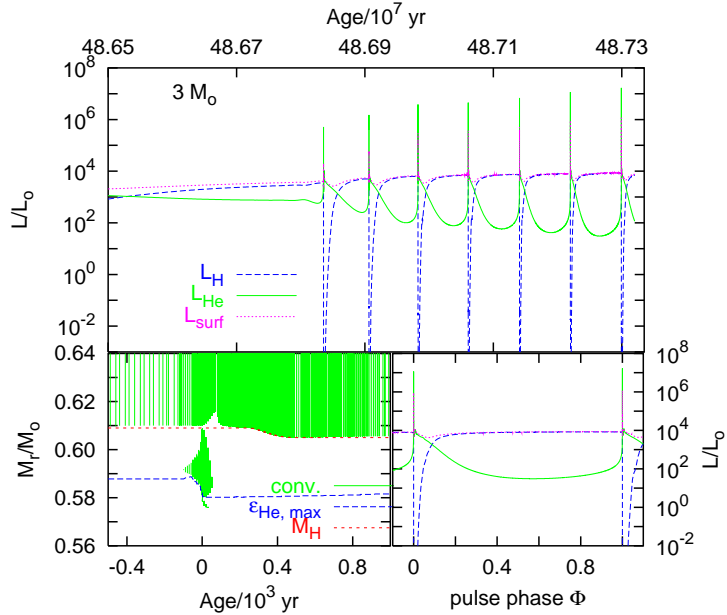}}
\caption[dup]{
Evolution of the surface luminosity and the luminosities of the H
and He burning shell during the first thermal pulses
of a 3 M$_{\odot}$ AGB sequence (top). The lower left panel gives the 
extension in mass of convective regions during the sixth pulse cycle. It
shows the formation of the pulse-driven convection zone and the following
dredge-up phase. Time is set to zero at maximum $L_{\rm He}$. 
$M_{\rm H}$ refers to  the mass of the H exhausted core, and 
$\varepsilon_{\rm He,max}$ to max.\ energy generation of He burning.
The lower right panel shows
the luminosity evolution as a function of the thermal pulse phase during
this cycle. Note that the left panel refers only to phases of
$\phi \sim 0.99$ to 0.01.
} \label{Ftp3m}
\end{figure}

On the upper AGB the helium burning shell becomes recurrently unstable
raising the so-called thermal pulses 
(Schwarzschild \& H\"arm 1965, Weigert 1966). During these
instabilities the luminosity of the He shell increases rapidly for a
short time of 100\,yr to $10^{6}$ to $10^{8}$\,L$_{\odot}$.
The huge amount of energy produced forces the development of a pulse-driven
convection zone which mixes products of He burning, i.e.\ carbon and oxygen,
into the intershell region.
Because the hydrogen shell is pushed concomitantly into cooler domains
hydrogen burning ceases temporarily allowing the envelope convection to proceed
downwards after the pulse,
to penetrate those intershell regions formerly enriched with
carbon (and oxygen) and to mix this material to the surface.
This 3$^{\rm rd}$ dredge up leads finally to the formation of carbon stars.
After the pulse hydrogen burning re-ignites and provides again the main
source of energy. Typically, a thermal-pulse cycle last
a few  $10^{3}$\,yr for large core masses ($\ga 0.8 $M$_{\odot}$) and  
$10^{4}$\,yr to $10^{5}$\,yr for smaller ones.
Fig.~\ref{Ftp3m} shows the evolution of the luminosities of the hydrogen und
helium burning shells during the thermal pulses
of a 3 M$_{\odot}$ AGB sequence.

\subsection{The role of overshoot}
The formation of carbon stars and its modelling by stellar evolution
calculations has been a long-standing problem.  
Early stellar evolution calculations predicted 
dredge up mostly for large core-masses, i.e. high luminosities,
whereas most of the (LMC) carbon stars  observed are found
at low luminosities and, therefore, are of lower mass
({\it carbon star mystery}, Iben 1981).
Moreover, various evolutionary calculations did not find dredge up at all
or differ in the efficiency obtained
(e.g.\ Vassiliadis \& Wood 1993, Bl\"ocker 1995a,
Forestini \& Charbonnel 1997).
The lack of optically bright carbon stars can be explained in 
terms of hot bottom burning and mass loss. HBB 
counteracts the carbon star formation.
However, since it ceases earlier than dredge up,
final dredge-up episodes may still produce carbon stars in the end
(Frost et al.\ 1998). Due to mass loss
such luminous carbon stars are expected to be
heavily enshrouded by dust hiding them from optical surveys.
Indeed, infrared observations of van Loon et al.\ (1999) show evidence
for the existence of luminous obscured (MC) carbon stars.
%
The failure to obtain efficient dredge up, in particular for
lower-mass AGB
models, is related to the only approximate description of stellar
convection and accordingly to the treatment of convective boundaries.
Another question intimately linked to this problem is how to produce
sufficient amounts of $^{13}$C in the interior as the neutron source for the
$s$-process nucleosynthesis.
Therefore, it is often concluded that mixing may take place
outside the formally convective boundaries
(Iben 1976, D'Antona \& Mazzitelli 1996,  Wood 1997).

To overcome these problems Herwig et al.\ (1997)
have considered diffusive overshoot for all convective boundaries
during the complete evolution leading to a considerable change in the models.
The overshoot prescription is based 
on the hydrodynamical calculations of Freytag et al. (1996) who 
showed that mixing takes place well beyond the classical
Schwarzschild border due to overshooting convective elements 
with an exponentially declining velocity field.
In the overshoot region the corresponding diffusion coefficient is given by
$D_{\rm os}=v_0\cdot H_{\rm p}\cdot\exp\frac{-2z}{f\cdot H_{\rm p}}$
with $v_0$: velocity of the convective elements immediately before the
Schwarzschild border; $z$: distance from the edge of the convective zone;
$f$: the overshoot efficiency parameter.
Within the stellar evolution calculations the  
efficiency parameter was chosen to be  $f=0.016$ as appropriate to match the
observed width of the main sequence.
This method provides for AGB stars a sufficient amount of dredge up
to form low-mass carbon stars as required by the observations.
Additionally it leads to the formation of $^{13}$C as a neutron source to
drive the $s$-process in these stars (see also Herwig et al.\ 1999a).

On the one hand, these calculations show that dredge up can easily be obtained
if some envelope overshoot is present to overcome the H/He discontinuity
by deepening the envelope convection.
On the other hand, overshoot leads also
to an enlargement of the pulse-driven convection zone and
to enhanced mixing of core material from deep layers below the He shell
to the intershell zone (``intershell dredge-up'') resulting in a 
considerable change of the intershell abundances.
After intershell dredge up
the abundances (mass fractions) of
(He,C,O) amount typically to (40,40,16)
instead of (70,25,2) as in non-overshoot sequences.
It is important to note that the total amount of dredge up depends mainly on
the strength of the former intershell dredge-up (Herwig et al.\ 1999a).
These modified intershell abundances are close to the 
observed surface abundances of Wolf-Rayet central stars 
(Koesterke, this volume) and PG\,1159 stars (Werner, this volume) 
and will turn out to be a  key ingredient for the modelling of these objects. 

\section{Evolution beyond the AGB}
\begin{figure}
\begin{minipage}{6cm}
\vspace*{1.5mm}
\epsfxsize=1.0\textwidth
\mbox{\epsffile{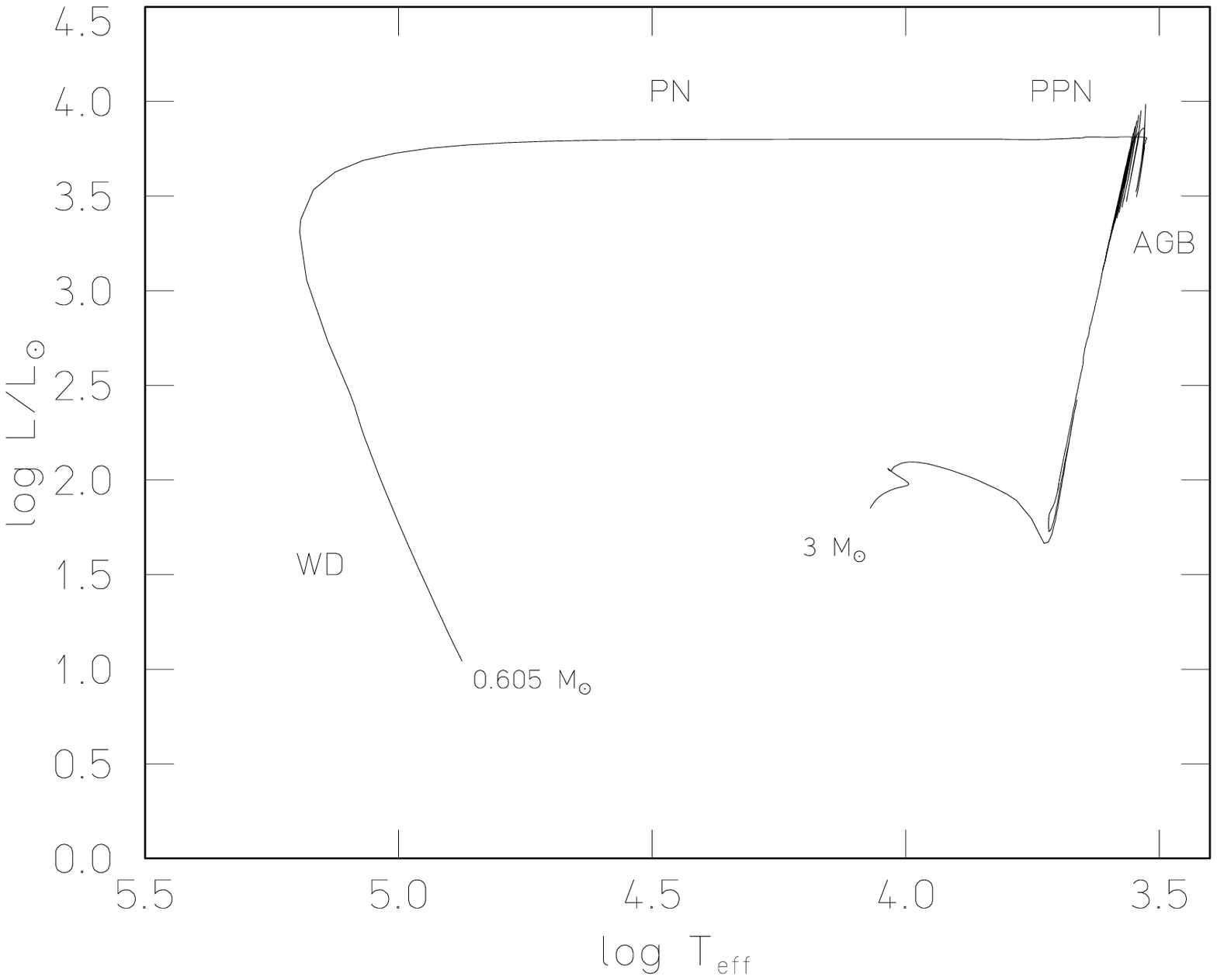}}
\end{minipage}
\begin{minipage}{6cm}
\hspace*{1.5cm}
\epsfxsize=0.83\textwidth
\mbox{\epsffile{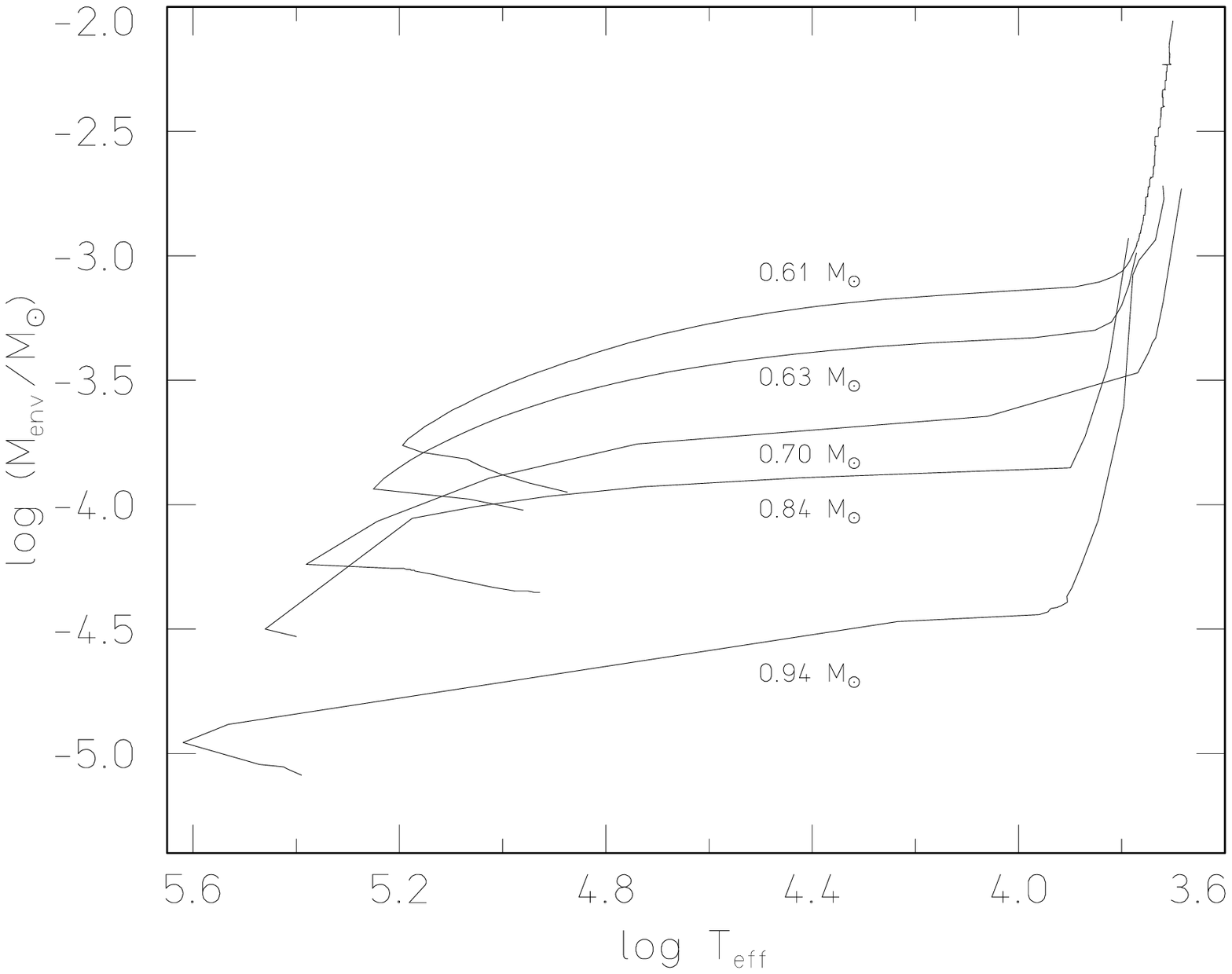}}
\end{minipage}
\vspace*{-1cm}
\caption{Left: Evolutionary path for an initially  3\,M$_{\odot}$ star 
 from the main sequence to the white-dwarf stage. 
 The last thermal pulses 
 at the tip of 
 the AGB are clearly visible. The model is burning hydrogen until the shell
 source gets extinct. The final mass is 0.605\,M$_{\odot}$.
 Right: Envelope mass vs. effective temperature for hydrogen burning
  post-AGB models (Bl\"ocker 1995b).
  The initial and remnant masses are
  (3\,M$_{\odot}$,0.61\,M$_{\odot}$),
  (3\,M$_{\odot}$,0.63\,M$_{\odot}$),
  (4\,M$_{\odot}$,0.70\,M$_{\odot}$),
  (5\,M$_{\odot}$,0.84\,M$_{\odot}$), and
  (7\,M$_{\odot}$,0.94\,M$_{\odot}$), resp. (from top to bottom).
}\label{Fmenvall}
\end{figure}

Mass loss terminates the AGB evolution
when the envelope mass is reduced to  
$\approx 10^{-2}\,$M$_{\odot}$. The star 
moves off the AGB (Sch\"onberner 1979) evolving towards the regime 
of central stars of planetary nebulae and finally reaches the stage of white
dwarfs. 
Fig.~\ref{Fmenvall} illustrates the evolution of an initially
3\,M$_{\odot}$ star from the main sequence to the white dwarf stage.
Recent reviews on the strucure and
evolution of central stars of planetary nebulae are given, e.g., by
Iben (1995) and Sch\"onberner and Bl\"ocker (1996).

%
%
\subsection{Mass Loss}
One important aspect of the transformation of AGB stars into white dwarfs
is the treatment of mass loss on the AGB and beyond. On the one hand, the AGB
mass-loss history determines the star's internal structure reached at the tip
of the AGB and therefore the fading speed  along the cooling part of the 
post-AGB evolution (Bl\"ocker 1995a,b).
On the other hand, the transition times from the AGB to
the central-star region depend sensitively on the mass-loss rates employed
beyond the AGB (Sch\"onberner 1983).
Observations indicate that mass-loss should decrease by order of magnitudes
during this transition phase (Perinotto 1989).
However,  up to now it is only poorly known how and 
at which temperature (range) this
strong decrease takes place.
The transition times should not be too long
since  the coolest post-AGB stars known have  
effective temperatures of about 5000\,K (Sch\"onberner \& Bl\"ocker 1993),
and kinematical ages of the youngest planetaries are only of the order of
1\,000 years. Fig.~\ref{Fmenvall} shows the envelope mass for different 
remnant masses as a function of the effective temperature. 

In the PN region mass-loss can be described
by the radiation-driven wind theory (Pauldrach et al. 1988). 
The respective mass-loss rates can be adapted like
$ \dot{M}_{\rm CPN} = 1.3 \cdot 10^{-15} \, L^{1.9}$ (Bl\"ocker 1995b),
leading to rates of $10^{-8}$ to $10^{-7} $M$_{\odot}$/yr for remants of
0.6 to 0.8\,M$_{\odot}$.
The corresponding influence on the  
evolutionary speed depends on the respective ratio of the burning rate
to mass-loss rate and is only important for massive remnants.
For H burning objects
the total crossing time of the lighter central stars is 
therefore uniquely given by the available fuel (i.e.\ the envelope mass) 
divided by the hydrogen luminosity. Because the envelope mass decreases 
and the luminosity increases 
with remnant mass, one obtains
typical crossing times (between 10\,000 K to the turn-around point in the 
HRD) of $\sim 100000$\,yr for 0.55 M$_{\odot}$, 
4000\,yr for 0.6 M$_{\odot}$, and 50\,yr for 0.94 M$_{\odot}$. 

When hydrogen burning cannot be sustained any longer because the 
envelope mass becomes too small, the surface luminosity must drop very
fast by at least one order of magnitude 
until it can be covered by gravothermal energy releases.
Helium burning is unimportant for most phase
angles and dies away as well. The fading time of AGB remnants down to, 
for instance, 100 L$_{\odot}$ is thus controlled by the gravothermal
energy release and neutrino energy losses. Both processes
depend on the thermomechanical structure of the core, and thus on 
the complete evolutionary history. 

\subsection{Final thermal pulses}
The evolution off and beyond the AGB depends also
on the thermal-pulse cycle phase
(fraction of the time span between two subsequent pulses)
$\phi$ with which the star moves off the AGB.
The post-AGB evolution is dominated
by helium burning for
$0 \le \phi \le 0.15$. 
For $0.15 \le \phi \le 0.3$ both nuclear shell sources contribute equal
luminosity fractions, for $0.3 \le \phi \le 1.0$ hydrogen burning determines
the nuclear energy production 
(Iben 1984). 
If the thermal-pulse cycle-phase
is sufficiently large, 
a last thermal pulse can occur
during the post-AGB evolution transforming a hydrogen burning into a helium
burning  model.
The flash forces the star to expand rapidly to Red Giant dimensions, and the 
remnant quickly evolves back to the AGB (``born again scenario''). There, it  
starts its post-AGB evolution again, but now as a helium burning object
(Iben 1984). The crossing timescale is now roughly three times larger than in 
the case of hydrogen burning objects. 
In principle we can distinguish three scenarios of final thermal
pulses relevant for Wolf-Rayet CSPNe (see also Herwig, this volume):
\begin{itemize}
\item[1.] The ``AGB final'' thermal pulse (AFTP),
   occurring immediately before the star moves off the AGB. In this case
   the envelope mass is already very small 
   ($\sim 10^{-2} $M$_{\odot}$, Fig.~\ref{Fmenvall}).
   If dredge up is able to operate even for such small envelope masses,
   a substantial fraction of the intershell region will be mixed with the
   tiny envelope leading to the dilution of hydrogen and enrichment
   with carbon and oxygen. However,
   standard evolutionary calculations predict a decreasing dredge-up 
   efficiency for low envelope masses (Wood 1981) and lack a sufficiently
   high oxygen abundance in the intershell region.
\item[2.] The ``late'' thermal pulse (LTP),
   occurring when the model evolves
   with roughly constant luminosity from the AGB towards the white dwarf
   domain. This kind of thermal pulse is similar to those experienced by
   AGB stars but the envelope mass is even smaller than for the AFTP
   ($\sim 10^{-4} $M$_{\odot}$, Fig.~\ref{Fmenvall}). 
   Evolutionary calculations without overshoot
   (e.g. Bl\"ocker \& Sch\"onberner 1997)
   predict only mild, if any, mixing. The envelope convection
   does not reach the layers enriched with carbon, and no
   3$^{\rm rd}$ dredge up occurs.
   Therefore, this scenario has often been considered in connection with 
   mass loss, which, however, cannot expose the hydrogen-free layers before
   effective temperatures of 100000\,K are reached (Iben 1984, Sch\"onberner
   \& Bl\"ocker 1992).
\item[3.] The ``very late'' thermal pulse (VLTP),
   occurring when the model is already
   on the  white dwarf cooling track, i.e.\ after the cessation of  H burning.
   In this case the pulse-driven convection zone of the He-burning shell
   may reach and penetrate the H shell causing a considerable or even
   total burning of hydrogen (Fujimoto 1977, Sch\"onberner 1979, 
   Iben 1984, Iben \& McDonald 1995). Due to mixing of the very tiny envelope
   ($\la 10^{-4} $M$_{\odot}$, Fig.~\ref{Fmenvall}) with substantial fractions
   of the 100 times more massive intershell region, the resulting surface
   abundances of carbon and oxygen are close to those of the intershell region.
   This scenario has been considered as the most promising one for 
   Wolf-Rayet CSPNe although it
   failed to match the observed oxygen abundances.
\end{itemize}

In summary one may conclude that within standard evolutionary calculations
neither the AFTP nor the LTP or VLTP appear to be well suited scenarios for 
Wolf-Rayet stars leaving these stars still enigmatic.
However, the consideration
of overshoot leads to a considerable change in the models making these
scenarios much more promising. 

\begin{figure}[t]
\centering
\epsfxsize=0.95\textwidth
\hspace*{-0.5cm}
\mbox{\epsffile[4 29 770 530]{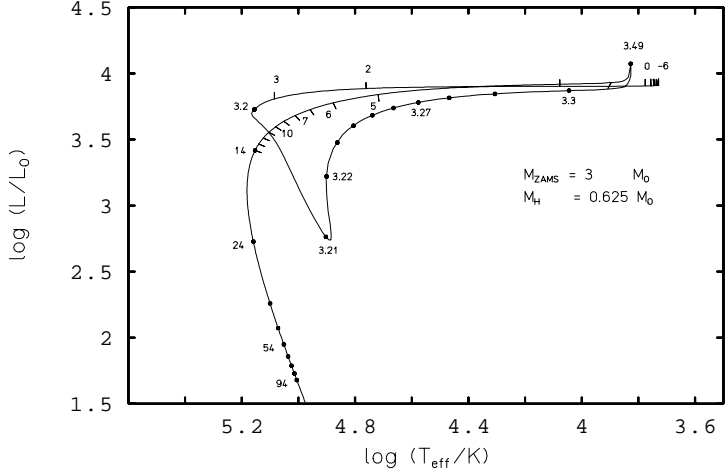}} 
\caption[heburn]{Evolution of a post-AGB model with 
 $M_{\rm ZAMS} = 3 $M$_{\odot}$ and $M_{\rm H} =  0.625 $M$_{\odot}$
 suffering from a LTP (Bl\"ocker 1995b). Age zero
 refers to a pulsational period of 50\,d.
 Time marks are in units of $10^{3}$ yrs. 
 } \label{Ffgsge}
\end{figure}

\subsection{AGB Final Thermal Pulse (AFTP)}
If overshoot is applied to all convective regions, AGB models
show efficient dredge up even for very low envelope masses (see Herwig, this
volume). Then, an AFTP can lead to both a considerable enrichment with carbon
and oxygen and to the dilution of hydrogen. The resulting surface abundances
depend on the actual envelope mass at which the AFTP occurs. For instance, 
Herwig (this volume) found for M$_{\rm env}= 4 \cdot 10^{-3} $M$_{\odot}$ 
(H,He,C,O)=(17,33,32,15) 
after the AFTP. The smaller M$_{\rm env}$, the closer are the final abundances
 to those observed in WR CSPNe. To obtain a sufficiently high probability 
for the AFTP to occur at very small envelope masses (i.e. at $\phi \approx 0$)
requires, however, a coupling of mass loss to the thermal pulse cycle.
Otherwise, the hydrogen abundance will be relatively high. This scenario
predicts small kinematical ages for the PNe of WR central stars which 
emerge here directly from the AGB.

\subsection{Late Thermal Pulse (LTP)}
The LTP, occurring for stars which leave the AGB with $\phi \ga 0.85$, 
have only been considered in connection with mass loss to explain the 
exotic surface abundances of WR CSPNe since dredge up is lacking.
In contrast to hydrogen burning objects where high mass loss leads
to a strong acceleration of the evolution limiting the removable mass, 
the longer evolutionary timescales of helium burning objects allow to 
expose the H-free layers, although only for $T_{\rm eff}$
in excess of  100000\,K. To strip off enough mass to expose even the deep
layers which match the observed carbon and oxygen abundances requires 
long lasting stellar winds. 
Although this scenario appears not to be applicable to WR CSPNe, it may 
have some relevance for 
the most exotic PG\,1159 star H\,1504. 
H\,1504 populates the cooling branch and shows a photosphere 
containing only carbon and oxygen with 50\% each (see Werner, this volume).
An exploratory study of Sch\"onberner \& Bl\"ocker (1992) showed that the 
photospheric parameters ($T_{\rm eff}$, $\log g$, surface abundances) 
of H\,1504 can be matched with a $0.84\,$M$_{\odot}$ LTP model suffering from a
constant mass-loss rate of $10^{-7}\,$M$_{\odot}$/yr until the position of 
H\,1504 in the HRD is reached. It is an open question
if such high mass-loss rates can be kept on the low luminosity part of the 
cooling branch. However, recent studies of Werner et al. (1995) 
give evidence that
hot white dwarfs may indeed suffer from such strong stellar winds.
\begin{figure}
\begin{center}
\hspace*{0.2cm}
\epsfxsize=0.6\textwidth
\epsfbox[53 61 556 566]{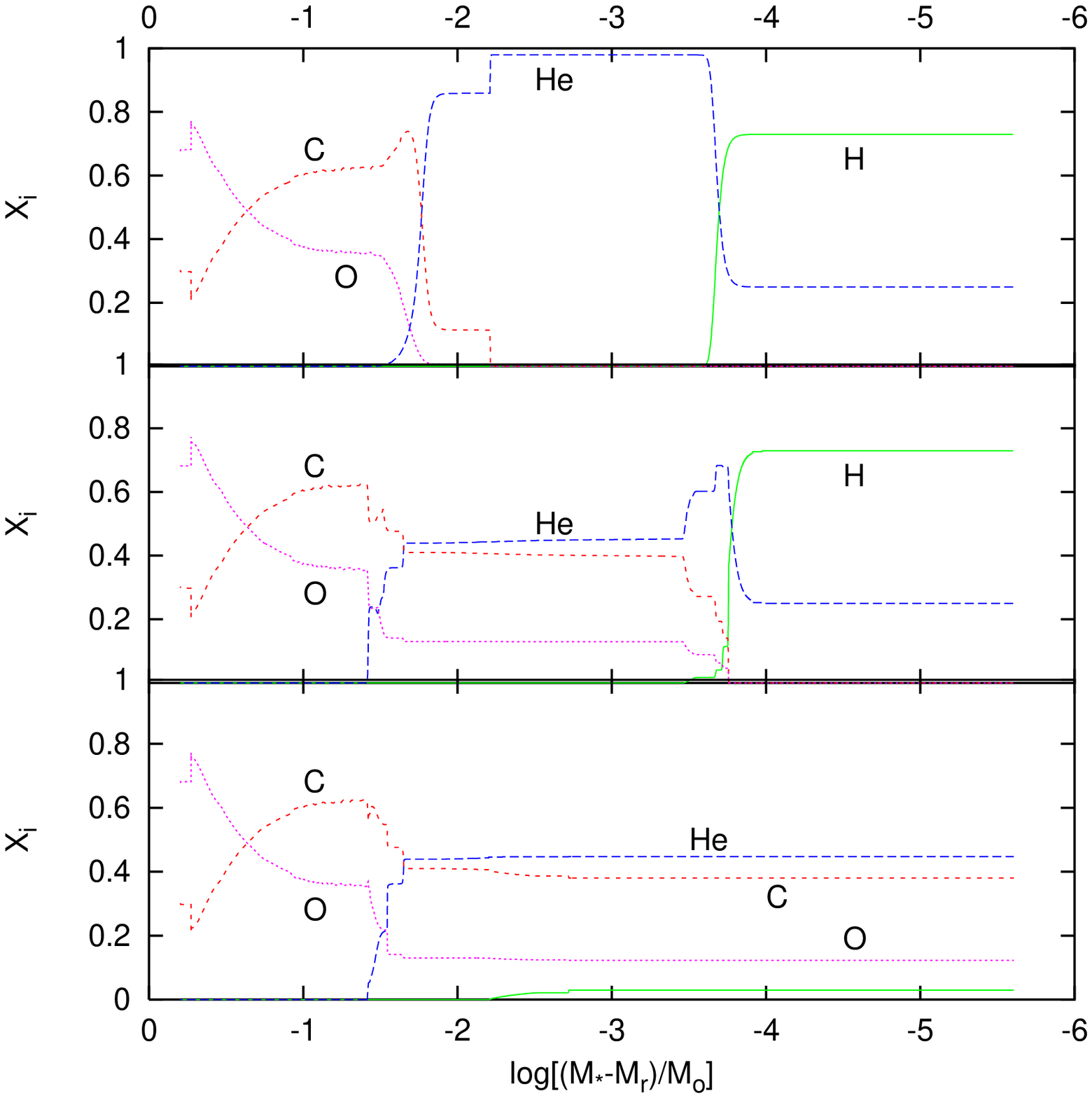}
\caption[heburn]{Abundances of H, He, C, and O as function of mass for a
post-AGB model of 0.625\,M$_{\odot}$.
The stellar
center is left and the surface is to the right. The top panel shows the
composition of the initial model calculated without overshoot on the AGB.
The middle panel refers to the point of time after the flash calculated with
enhanced overshoot efficiency as required to generate appropriate intershell
abundances. The situation after the dredge up is shown in the lower panel.
 } \label{Fmlay}
\end{center}
%
\begin{center}
\hspace*{0.4cm}
\epsfxsize=0.7\textwidth
\epsfbox[50 58 557 414]{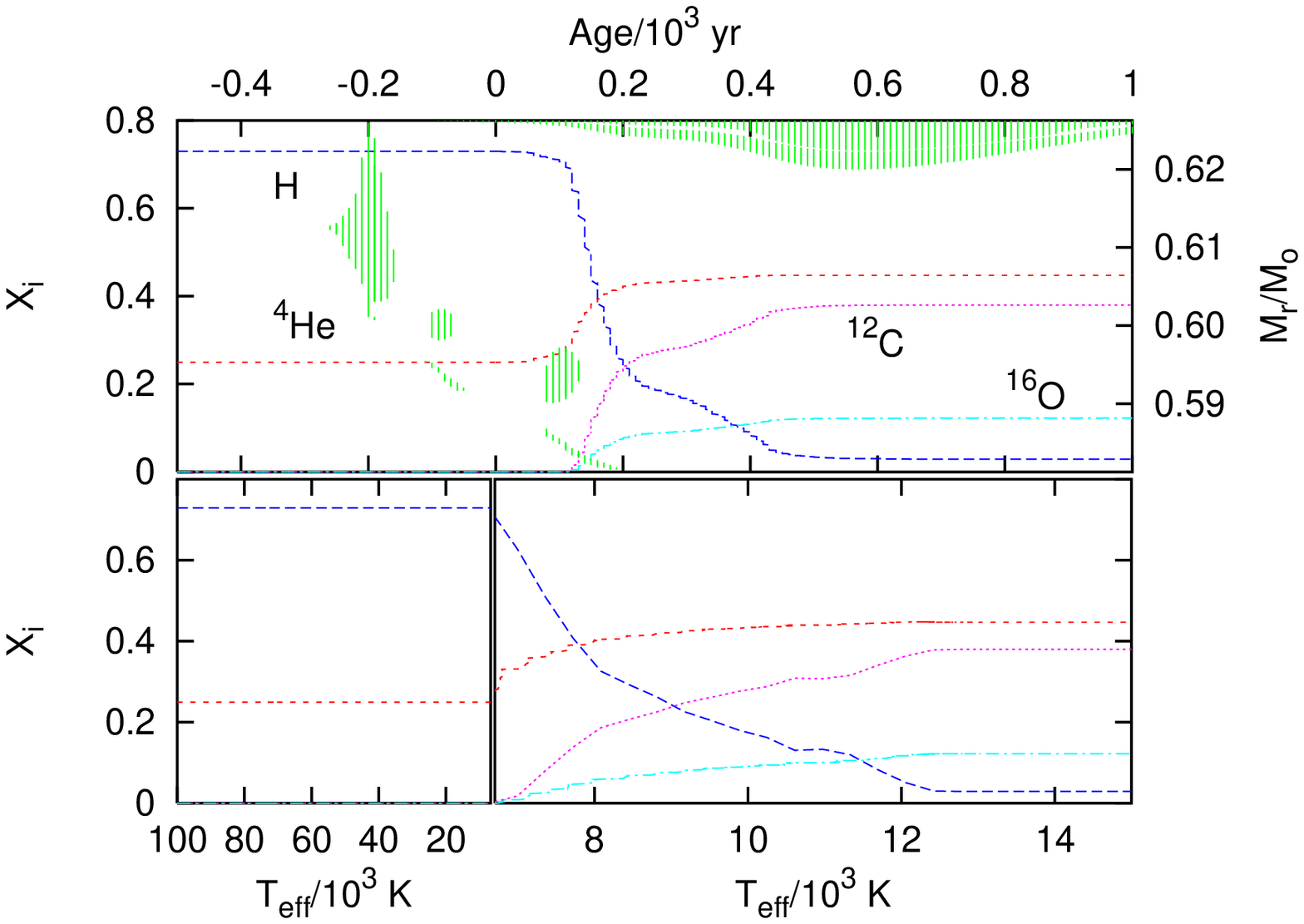}
\caption[heburn]{
Evolution of the surface abundances of H, He, C, and O and extension of
convective regions for the  0.625\,M$_{\odot}$ model during the flash and 
dredge up (top). Shaded regions denote convective regions. Time is set to
zero at minimum effective temperature. The lower panel gives the corresponding
evolution for the abundances as function of the effective temperature. 
} \label{Fltpkipp}
\end{center}
\end{figure}

So far, stellar evolution calculations did not show any dredge up mixing 
carbon to the surface within the LTP scenario. However, if overshoot is 
considered, efficient dredge up operates even for 
$M_{\rm env} \la 10^{-4} $M$_{\odot}$. These new models show for the first
time that the LTP scenario may serve as one possible channel for the WR CSPNe.
We note that, as in the case of the AFTP, overshoot applied to the pulse-driven
convection zone is essential for the resulting dredge-up efficiency.
We have applied the overshoot scheme of Herwig et al.\ (1997) to 
the 
post-AGB models of Bl\"ocker (1995b) which are based on an
AGB evolution without overshoot. Herwig et al. (1999a) have shown that the 
intershell abundances of carbon and oxyen first increase and then level out 
after a series of thermal pulses for overshoot AGB models. 
To account 
for these typical intershell abundances, we re-calculated the LTP of 
Bl\"ocker's (1995b) 0.625\,M$_{\odot}$ model with an enhanced overshoot
efficiency (f=0.064) for the pulse-driven convection zone 
leading to (He,C,O)=(45,40,13) in the intershell region.
For the remaining evolution the standard efficiency parameter of
f=0.016 was used. This method provides a simulation close to an object where 
overshoot have been applied during the whole preceding AGB evolution.
Fig.~\ref{Fmlay} shows the abundances of H, He, C and O 
for the initial model and immediately after the flash.

After the flash (and two subpulses, see Fig.~\ref{Fltpkipp}) the model evolves
towards the AGB domain. At minimum effective temperature ($\approx 6700$\,K)
dredge up sets in 
and continues until the star has reheated to $\approx 12000$\,K.
Hydrogen is diluted to 3\% and the final surface abundances of He, C and O 
are close to those of the intershell region, viz.\ (45,38,12). Extension of 
convective regions and abundances are illustrated in Fig.~\ref{Fltpkipp} as 
function of age and effective temperature, resp.
We obtain surface abundances as observed in 
WR CSPNe (cf.\ Koesterke, this volume). The kinematical age of the PN amount 
to a few thousand years (see Fig.~\ref{Ffgsge}). 

This scenario applies to FG\,Sge as well that suffered from a LTP roughly 
100 yr ago. FG\,Sge seems to have already reached its minimum effective
temperature and to reheat now again (Kipper 1996). Its mass can be estimated
to be close to 0.6\,M$_{\odot}$ and its surface stays hydrogen-rich during 
its evolution back to the AGB 
(excluding a VLTP, Bl\"ocker \& Sch\"onberner 1997). However, recently 
evidence is growing that FG\,Sge becomes hydrogen-deficient during its 
reheating (Gonzalez et al.\ 1998). This is hard to explain within the
standard LTP scenario (no overshoot)
but is completely in line with the predictions of the new models. 

\subsection{Very Late Thermal Pulse (VLTP)}
The VLTP is different from the AFTP and LTP since it causes a burning (and
mixing) of the envelope requiring more demanding numerics. The only model
available until recently was that of Iben \& McDonald (1995)
which shows strong hydrogen 
deficiency after the flash but too low oxygen surface abundances as to 
account for the WR\,CSPNe. Herwig et al. (1999b) presented the 
first VLTP model which achieves a general agreement with the abundance
patterns of WR\,CSPNe due to the consideration of overshoot.
Again, the existence of intershell dredge-up providing 
intershell abundances close to those observed in photospheres
of WR CSPNe appeared to be crucial to match the 
observations (see also Herwig, this volume).

The VLTP occurs on the cooling track when H burning is already off. In this
instance, the pulse-driven convection zone can reach and penetrate the H-rich
envelope due to the lack of an entropy barrier (Iben 1976) built up by 
H burning. Then, protons are ingested into the hot, carbon-rich intershell 
region and are captured via $^{12}$C(p,$\gamma$)$^{13}$N. Convective turnover
timescales and nuclear timescales become comparable during this phase, and the 
protons are burnt on their way towards the interior. Accordingly, a 
simultaneous treatment of mixing and burning is essential to treat this phase 
of evolution correctly.  The ingestion of protons finally raises a H flash
and the energy released by this flash leads to a splitting of the convection 
into a upper zone powered by H burning and a lower one powered by He burning. 
The upper convection zone is, however, short-lived because the available 
hydrogen in the envelope is quickly consumed. Finally, the star becomes 
hydrogen-free and exposes its intershell abundances at the surface. 

\begin{figure}
\begin{minipage}{7.0cm}
\epsfxsize=1.0\textwidth
\epsfbox{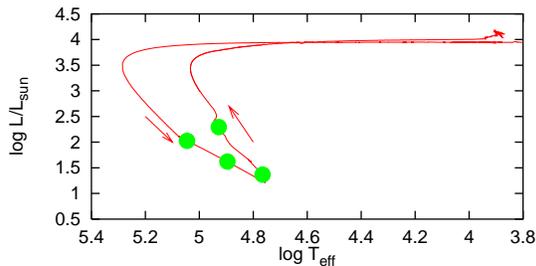}
\end{minipage}
\begin{minipage}{5.0cm}
\caption[heburn]{Evolutionary track of a  0.604\,M$_{\odot}$ post-AGB model 
 ($M_{\rm ZAMS} = 2 $M$_{\odot}$)
 suffering from a VLTP (Herwig et al.\ 1999b). The dots refer to the
 formation of the pulse-driven convection zone, 
 to the ingestion of protons, to the peak
 of the induced H flash, and to the moment when all hydrogen is burnt
 leaving a hydrogen-free surface.
 } \label{Fvlt}
\end{minipage}
\end{figure}

Fig.~\ref{Fvlt} shows the evolutionary track of the 0.604\,M$_{\odot}$ 
model of 
Herwig et al.\ (1999b) which is based on a 2\,M$_{\odot}$ overshoot sequence.
The dots on the track denote the establishment of the 
pulse-driven convection zone, the ingestion of protons, 
the peak of the H flash,
and the moment when the surface becomes H-free. The resulting surface 
abundances are (He,C,O)=(38,36,17). Among the remaining elements 
is Ne with 3.5\%. Note that the burning of the envelope takes place at 
high effective temperatures and that no hydrogen survives. The kinematical
age of the surrounding PN is relatively high within the VLTP scenario since 
the star has first to fade along the cooling branch down to a few 
100\,L$_{\odot}$ before the onset of the flash. 
For 0.6 M$_{\odot}$ one 
obtains typically $t \ga 20000$\,yr. 
%
%
%
%

\section{Concluding remarks}
If overshoot is applied to AGB models, 
intershell dredge-up provides intershell abundances
close to the surface abundances of Wolf-Rayet CSPNe and PG\,1159 stars.
These intershell abundances can be mixed to the surface during the AFTP, LTP
or VLTP scenario. In contrast to standard evolutionary calculations, 
overshoot models do show dredge up for very low envelope masses 
and efficient dredge up was found even during the post-AGB stage. 
All scenarios lead to hydrogen-deficient post-AGB stars with 
carbon and oxygen abundances as observed for Wolf-Rayet stars.
Hydrogen is either diluted by dredge up (AFTP, LTP) or completely 
burnt (VLTP). The AFTP leads to a relatively high hydrogen abundance
($\ga 15\%$) whereas the LTP gives only a few percent ($\la 3\%$).
The kinematical ages of the planetary nebulae are relatively high for the VLTP 
and are much lower for the AFTP and LTP scenarios.
The variety of observations requires most likely all of these scenarios.
Many objects have only very low hydrogen abundances, if any, favoring the 
LTP and VLTP. On the other hand, several Wolf-Rayet central stars are 
surrounded by young  planetary nebulae (Tylenda 1996) and circumstellar shells 
(Waters et al.\ 1998) strengthening the AFTP and LTP. 
Within the current models roughly 20 to 25\% of stars moving off the AGB 
can be expected to become hydrogen-deficient.

\subsection*{Acknowledgements} 
\vspace*{-1.5ex}
I would like to thank F.\ Herwig, T.\ Driebe, R.\ Waters and K.\ Werner 
for many valuable discussions.

\end{document}